# Integrated Graphene Patch Antenna for Communications at THz Frequencies

Elana P. de Santana[1], Anna K. Wigger[1], Zhenxing Wang[2], Kun-Ta Wang[2], Max Lemme[2], Sergi Abadal[3] and Peter Haring Bolívar[1]

[1]Institute for High Frequency and Quantum Electronics, University of Siegen, 57076 Siegen, Germany
[2]AMO GmbH, 52074 Aachen, Germany
[3]NaNoNetworking Center in Catalunya, Technical University of Catalonia, 08034 Barcelona, Spain

*Abstract*— Graphene is an attractive material for communications in the THz range due to its ability to support surface plasmon polaritons. This enables a graphene antenna to be smaller in size than its metallic counterpart. In addition, the possibility to control the graphene conductivity during operation by an applied bias leads to the tunability of the resonant frequency of graphene antennas. Graphene-based antennas integrated into transceivers working at THz frequencies may lead to faster and more efficient devices. In this work, we design and simulate a graphene patch antenna that can be integrated into transceivers by through-substrate vias. The tuning of the resonant frequency is also studied by simulations.

## I. INTRODUCTION

SURFACE plasmon polariton (SPP) is a type of surface wave traveling along with a metal-dielectric interface with a smaller wavelength compared to the free-space wavelength at the same frequency [1]. In the THz range, a system consisting of dielectric-graphene-dielectric supports this type of surface wave turning a graphene-based antenna ideal for communications in the THz band since it is smaller than a metallic antenna resonating at the same frequency. The fact that the conductivity of the graphene can be tuned electrically by an applied bias or by chemical doping [2] and that the graphene conductivity, which can be described by the Kubo formula [3], is directly related to the graphene SPP wavelength allows the resonant frequency of a graphene antenna to be tunable without changing its physical dimensions.

Smaller antennas facilitate their integration directly with the electronic components to be assembled into the transceiver. We propose a graphene antenna to be integrated into the transceiver by using vias connecting the back-metal plane to the top ground pads (Figure 1). The SiGe or CMOS transceiver can be flipped and bonded to the signal and ground pads of the antenna. The performance of the antenna and its tuning possibility are evaluated based on electromagnetic simulations.

## II. RESULTS

The graphene patch antenna, represented in Figure 1, is designed using the standard antenna equations that can be found in [4] to resonate at the J-band of the spectrum, most precisely at 280 GHz, leading to lateral dimensions of 355 µm x 260 µm. A metal through-substrate via (TSV) connects the ground pads to the back-metal plane. This is assumed to be an aluminum wire of a 5 µm radius. The substrate material is chosen to be polyimide (ε = 3.5, δ = 0.0027) with 50 µm thickness with a back-metal plane made of aluminum. We use CST as the simulation software with the transient solver to acquire the resonant frequency and gain of the antenna in the range of 220 – 325 GHz and used a waveguide port to excite the antenna.

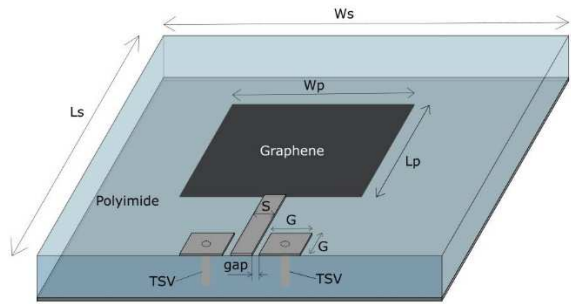

**Fig. 1.** Design of the graphene patch antenna fed by a coplanar waveguide used for simulations showing the vias connecting the top and bottom ground metals with a radius R.

Other design specifications can be found in Table 1.

TABLE I
GRAPHENE ANTENNA DESIGN PARAMETERS

| Parameter | Value (µm) |
|---|---|
| Patch Width ($W_p$) | 355 |
| Patch Length ($L_p$) | 262 |
| Substrate Width ($W_s$) | 2*$W_p$ |
| Substrate Lenth ($L_s$) | 2*$L_p$ |
| Substrate Thickness ($t_s$) | 50 |
| Signal Pad Width (S) | 40 |
| Ground Pad Width (G) | 50 |
| Gap | 5 |
| TSV Radius (R) | 5 |

The graphene patch is simulated as a 2D material with surface conductivity, $\sigma$, defined by the Kubo formula, as described in Equation (1), where $e$ is the electron charge, $\tau$ is the relaxation time of the graphene, $\hbar$ is the reduced Planck's constant, $K_b$ is the Boltzmann constant, $\mu_c$ is the graphene Fermi level and $\omega$ is frequency.

$$\sigma(\omega) = \frac{2e^2}{\pi\hbar}\frac{k_B T}{\hbar}\ln\left[2\cosh\left[\frac{E_F}{2k_B T}\right]\right]\frac{i}{\omega + i\,\tau^{-1}} \qquad (1)$$

Comparison of the simulated S11 results for a metal patch antenna and graphene patch antennas can be found in figures 2 and 3. For all graphene patch antenna simulations, the resonant frequency is smaller than that of a metal antenna with the same dimensions.

It is possible to tune the resonant frequency of the graphene antenna by changing the Fermi level of the graphene patch, cf. Figure 2. This is based on the change of carrier density, which can be done by applying a voltage to the graphene patch. By changing the graphene Fermi level from 0.3 eV to 1.2 eV, the resonant frequency of the antenna is shifted from 225 GHz to 263 GHz. But while the graphene antenna with 0.3 eV reflects most of its received signal, the return loss is improved for the higher graphene Fermi levels, leading to more efficient antennas.

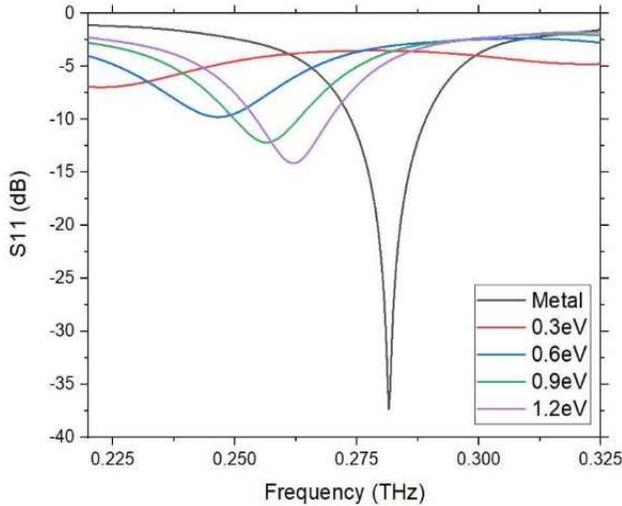

**Fig. 2.** Return loss comparison between a metal antenna and a graphene antenna with 1.2 ps of relaxation time and Fermi level varying from 0.3 eV to 1.2 eV

In addition, graphene antennas with higher relaxation times

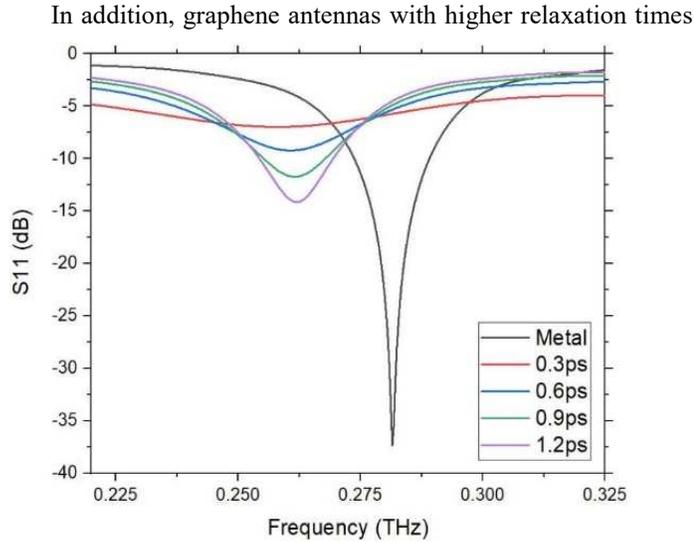

**Fig. 3** Return loss comparison between a metal antenna and a graphene antenna with 1.2 eV of Fermi level and relaxation time varying from 0.3 ps to 1.2 ps

show an improved performance (Figure 3). The higher the relaxation time of the graphene patch, the more power is received - and transmitted - by the antenna. The graphene relaxation time is related to the graphene quality and is affected by the graphene growing process, transfer to the target substrate and patterning.

Carrier mobility is another parameter used to identify a high-quality graphene. Graphene carrier mobility, $\mu$, can be calculated using Equation (2), where $v_F$ is the Fermi velocity in graphene with value $10^6$ m/s [2].

$$\mu = \frac{\tau e v_F^2}{E_f} \quad (2)$$

High carrier mobilities, e.g. 10.000 cm$^2$/Vs for graphene with 0.9 eV, are crucial for a graphene patch to behave as a good antenna (less than -10 dB of return loss and positive gain). Previous works have shown the same high mobility requirement for a dipole graphene antenna [5]. Nevertheless, low graphene mobilities can be compensated to a certain extent by higher Fermi levels.

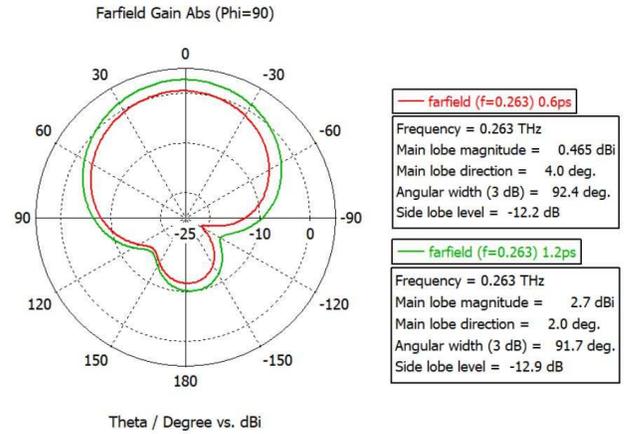

**Fig. 4** Simulated gain of a graphene antenna with a Fermi level of 1.2 eV and 0.6 ps (red) and 1.2 ps (green) of relaxation time.

Simulation results show that for lower relaxation times, the far-field gain of the antenna is decreased. The emission pattern (Figure 4) of a graphene antenna of 1.2 eV and 1.2 ps resonating at 263 GHz with a bandwidth of 19 GHz, here considered as the frequency range where return loss is smaller than -10 dB, shows that the main radiation lobe of this antenna is directed to the 2° with a gain of 2.7 dBi, while for a lower mobility antenna with 0.6 ps of relaxation time the main lobe direction is to the 4° with a gain of 0.4 dBi.

### III. CONCLUSION

The TSV permits the antenna to be connected to the transceiver die providing the RF signal to be emitted. The polyimide substrate provides not only good working conditions for the antenna but also provides the isolation required between the antenna and the RF circuit. The simulation results show that the graphene patch antenna with 1.2 eV of Fermi level presents a resonance frequency 6% lower compared to a metal antenna of the same dimensions. Consequently, this graphene antenna to resonate at the same frequency as the metal antenna will have dimensions of 355 μm x 220 μm, being 15% smaller in area reducing total chip size. In resume, our simulations show that, given an appropriate graphene quality, our graphene antenna design is well suitable for future integration of THz antennas with transceivers.

Acknowledgments: This project has received funding from the European Union's Horizon 2020 research and innovation program under grant agreement N° 863337.

### REFERENCES

[1] T. Low and P. Avouris, "Graphene Plasmonics for terahertz to Mid-infrared Applications," *ACS Nano,* vol. 8, no. 2, pp. 1086-1101, January 2014.
[2] Y. Yu, Y. Zhao, S. Ryu, L. E. Brus, K. S. Kim and P. Kim "Tuning the Graphene Work Function by Electric Field Effect," *Nano Letters*, vol. 12, n. 10, pp. 3430-3434, August 2010.
[3] M. Tamagnone, J. S. Gómez-Días, J.R. Mosig, and J. Perruisseau-Carrier, "Analysis and Design of Terahertz Antennas based on Plasmonic Resonant Graphene Sheets," *Journal of Applied Physics*, vol. 112, no. 114915, pp. 1-4, December 2012.
[4] C. A. Balanis, *Antenna Theory: Analysis and Design*, 4[th] ed.Hoboken, NJ: Wiley, 2016
[5] C. Suessmeier, S. Abadal, D. Stock, S. Schaeffer, and others, "Material-Dependencies of the THz Emission from Plasmonic Graphene-based Photoconductive Antenna Structures," *2017 42$^{nd}$ International Conference of Infrared, Millimeter, and Terahertz Waves (IRMMW-THz)*, pp. 1-2, 2017.